\documentclass[12pt]{iopart}

\pdfoutput=1
\usepackage{iopams}  
\usepackage[T1]{fontenc}
\usepackage[latin9]{inputenc}
\usepackage{setstack}
\usepackage{graphicx}
\usepackage{subfigure}

\begin{document}


\letter{Three-fold way to extinction in populations of cyclically competing species}
\author{S Rulands $^1$, T Reichenbach$^2$ and E Frey $^1$}

\address{$^1$ Arnold Sommerfeld Center for Theoretical Physics (ASC) and Center for NanoScience 
(CeNS), LMU M\"unchen, Theresienstra{\ss}e 37, 80333 M\"unchen, 
Germany}
\address{$^2$ Howard Hughes Medical Institute and Laboratory of Sensory Neuroscience, The Rockefeller University, 
1230 York Avenue, New York, NY 10065-6399}
\ead{frey@lmu.de}
\begin{abstract}
Species extinction occurs regularly and unavoidably in ecological systems. The time scales for extinction can broadly vary and inform on the ecosystem's stability. We study the 
spatio-temporal extinction dynamics  of a paradigmatic population model where three species exhibit cyclic competition. The cyclic dynamics reflects the non-equilibrium nature of the species interactions. While previous work focusses on the coarsening process as a mechanism that drives the system to extinction, we found that unexpectedly the dynamics to extinction is much richer. We observed three different types of dynamics. In addition to coarsening, in the evolutionary relevant limit of large times, oscillating traveling waves and heteroclinic orbits play a dominant role. The weight of the different processes depends on the degree of mixing and the system size. By analytical arguments and extensive numerical simulations we provide the full characteristics of scenarios leading to extinction in one of the most surprising models of ecology.

\end{abstract}

\pacs{87.23.Cc, 05.40.-a, 02.50.Ey}
\vspace{2pc}
\noindent{\it Keywords}: Phase transitions into absorbing states (Theory), Population dynamics (Theory), Stochastic processes, Coarsening processes (Theory)

\maketitle
\section{Introduction}
Stochastic many-particle systems provide a testing ground for non-equilibrium dynamics. In nature, 
systems frequently evolve away from equilibrium and then relax to an 
equilibrium  steady state. Understanding the relaxation process is a central topic in non-equilibrium physics. 
Near-equilibrium fluctuations are governed by the same laws that hold 
in steady state and the transient is typically an exponential decay. Many systems, however, comprise 
absorbing states, which can be reached but never be left by the dynamics. In 
this case no fluctuations are present in the steady states. Such systems arise in a broad variety of problems, e.g. 
physics, chemistry or epidemics ~\cite{Hinrichsen:2000p5571}. Much 
effort has been spent on the investigation of simple, diffusion-limited chemical reactions where the decay to equilibrium can obey power laws ~\cite{Taeuber-2005}.

Understanding transitions into absorbing states is not only fundamental for non-equilibrium physics, but is also highly relevant for ecology. Here, absorbing states correspond to 
the extinction of species. Another characteristic feature of ecological systems are cyclic interactions. As a classic example, the work of Lotka and Volterra describes the dynamics 
of fish populations in the adriatic as persistent oscillations due to predator-prey interactions. Other examples include coral reef invertebrates \cite{Jackson-1975}, rodents in the high arctic tundra in Greenland \cite{Gilg-2002}, cyclic competition between different mating strategies of lizards \cite{sinervo-1996-340} and chemical warfare of \emph{Escherichia coli} bacteria under laboratory conditions ~\cite{kerr-2002-418}.

Recent work has investigated cyclic competition in one-dimensional systems with no or only weak diffusion of the reacting agents~\cite{Tainaka-1988,Frachebourg:1996p5895,frachebourg-1996-54,Taeuber2010}. Coarse-graining of temporally growing and annihilating domains has been identified as the mechanism that eventually leads to species extinction. However, individual's mobility may be significant and alter this picture qualitatively.

In this article, we investigate the spatio-temporal dynamics of extinction in a paradigmatic model of three species in cyclic competition. Individuals are positioned on a one-dimensional lattice and are equipped with fast mobility that leads to effective diffusion. The system possesses absorbing states in 
the form of extinction of two of the three species and, because of fluctuations, the dynamics eventually comes to rest there. However, the time-scales until extinction occurs 
provide information on the stability of species diversity~\cite{cremer-2009-11}. We identify three distinct types of dynamics that lead to extinction. These types of dynamics arise 
from the possible influences that intrinsic fluctuations can have on the coarsening process and on the traveling waves that the cyclic dynamics induces. The different dynamics 
lead to characteristic dependences of the extinction-time probability on the elapsed time $t$ and the system size $N$. We provide semi-phenomenological arguments that quantify 
the functional form and the scaling behaviour of the extinction-time probability. These arguments yield 
information on the emergence and characteristics of the different types of 
dynamics.

\section{The model}

Consider a stochastic, spatial variant of the May-Leonard
model which serves as a prototype for cyclic, rock-paper-scissors-like species interactions. Three
species $A, B, C$ compete with each other in a cyclic manner, at rate $\sigma$, and reproduce at rate $\mu$ upon availability of empty space $\emptyset$:

\begin{eqnarray}
AB\stackrel{\sigma}{\rightarrow}A\emptyset, & BC\stackrel{\sigma}{\rightarrow}B\emptyset, & CA\stackrel
{\sigma}{\rightarrow}C\emptyset,\cr
A\emptyset\stackrel{\mu}{\rightarrow}AA, & B\emptyset\stackrel{\mu}{\rightarrow}BB, & C\emptyset\stackrel
{\mu}{\rightarrow}CC.
\label{eq:react}
\end{eqnarray}

For increasingly large populations intrinsic fluctuations eventually become negligible. If in addition spatial structure is absent, \emph{i.e.}, if every individual can interact with every 
other in the population at equal probability, the population dynamics is aptly described by deterministic rate equations for the densities
$\vec{s}=(a,b,c)$ of the species $A, B$ and $C$:
\begin{equation}
\partial_{t}s_{i}=s_{i}\left[\mu\left(1-\rho\right)-\sigma s_{i+2}\right], \quad\mbox{ for }i\in\{1,2,3\}.
\label{eq:re}
\end{equation}
Hereby the indices are understood as modulo $3$ and $\rho=a+b+c$ denotes the total
density. May and Leonard showed that these equations possess $4$ absorbing
fixed points, corresponding to the survival of one of the species
and to an empty system~\cite{May:1975p197}. Furthermore  a reactive
fixed point $s^{*}=\frac{\mu}{\sigma+3\mu}(1,1,1)$ exists that represents coexistence of all three species. Linear stability analysis  shows that $s^{*}$ is unstable. The absorbing steady states that correspond to  extinction, $(1,0,0)$, $(0,1,0)$ and $(0,0,1)$, are heteroclinic
points. The Lyapunov function $\mathcal{L}=abc/\rho^{3}$ demonstrates that the trajectories of the deterministic equations~(\ref{eq:re}), when initially close to the reactive fixed point,  spiral outward on an invariant manifold. On this manifold the trajectories then approach the boundary of the phase space and form heteroclinic cycles, converging to the boundary and the absorbing states  without ever reaching them.

However, intrinsic noise from finite-system sizes~\cite{claussen-2008-100, boland-2009-79} and spatial correlations alter the above behaviour~\cite{Durrett:1998p203, szabo-2007-446,abta-2007-98,PhysRevE.78.031906}. While fluctuations ultimately drive the system into one of the absorbing fixed points~\cite{parker-2009-80}, the formation of spatial patterns can substantially delay extinction and promote species coexistence~\cite{szabo-2007-446,Reichenbach2007,efimov-2008-78}. The resulting spatio-temporal dynamics of extinction is nontrivial and highly interesting.
%

\section{Numerical results} 
We consider a one-dimensional lattice of $L$ sites with periodic boundary conditions. Each lattice site hosts a fixed number $M$ of individuals $A,B,C$ and empty spaces $\emptyset$, such that the concentrations in the rate equations (\ref{eq:re}) are given by the number of particles of a specified type divided by $M$. $M$ may hence be viewed as the carrying capacity of a lattice site. The reactions~(\ref{eq:react}) occur between individuals on the same lattice site. Individuals may change place with another individual or an empty space on a neighbouring lattice site at rate $\epsilon$. 
In order to keep the length scale, \emph{i.e.} characteristic length scale for diffusion, fixed when changing the lattice spacing $L^{-1}$ we have to rescale $\epsilon$ appropriately. In the continuum limit, where (\ref{eq:re}) holds, the exchange processes therefore lead to an effective diffusion of individuals at a diffusion constant $D\equiv \epsilon L^{-2}$  and thus to coupling between the lattice sites.
In our simulations  we implemented a continuous-time Markov process with sequential updating. At each
simulation step an individual is randomly chosen. It then either reacts with
a randomly chosen individual of the same site or changes place
with a randomly chosen individual of the two neighbouring stacks,
at  probabilities corresponding to the rates $\sigma$, $\mu$ and
$\epsilon$.

The population model introduced above possesses a net system size of $N\equiv ML$ which plays the role of an overall carrying capacity. For large enough $M$ and $L$ the intrinsic fluctuations have a strength proportional to the inverse square-root of $N$~\cite{Gardiner2004}. Different equivalent ways therefore exist for performing the thermodynamic limit,  \emph{e.g.}, increasing the number $M$ of individuals per lattice site, while keeping the lattice size $L$ fixed or increasing the lattice size $L$, keeping $M$ fixed. Because for large $L$ a huge amount of exchange processes takes place between the reactions, requiring long computation time, we performed the thermodynamic limit in $M\rightarrow\infty$ and kept $L=100$ fixed. The insensitivity of the results to the choice of the limit is supported by recent studies~\cite{Taeuber2010}. $L$ was chosen sufficiently large, such that $L^{-1}$ was much smaller than the correlation length.

We here consider the case of equal reproduction and selection rates $\mu=\sigma=1$. Similar behaviour can be expected for $\mu\neq\sigma$, when $D$ is rescaled appropriately \cite{Reichenbach2007}, and species dependent interaction rates \cite{Taeuber2010,PhysRevE.81.021917}. We chose a random initial configuration in which the density of the species approximately equals the ones of the internal fixed point $s^{*}$ of the rate equations~(\ref{eq:re}). 
\begin{figure*}[t]
\begin{center}
\includegraphics[width=0.3\textwidth]{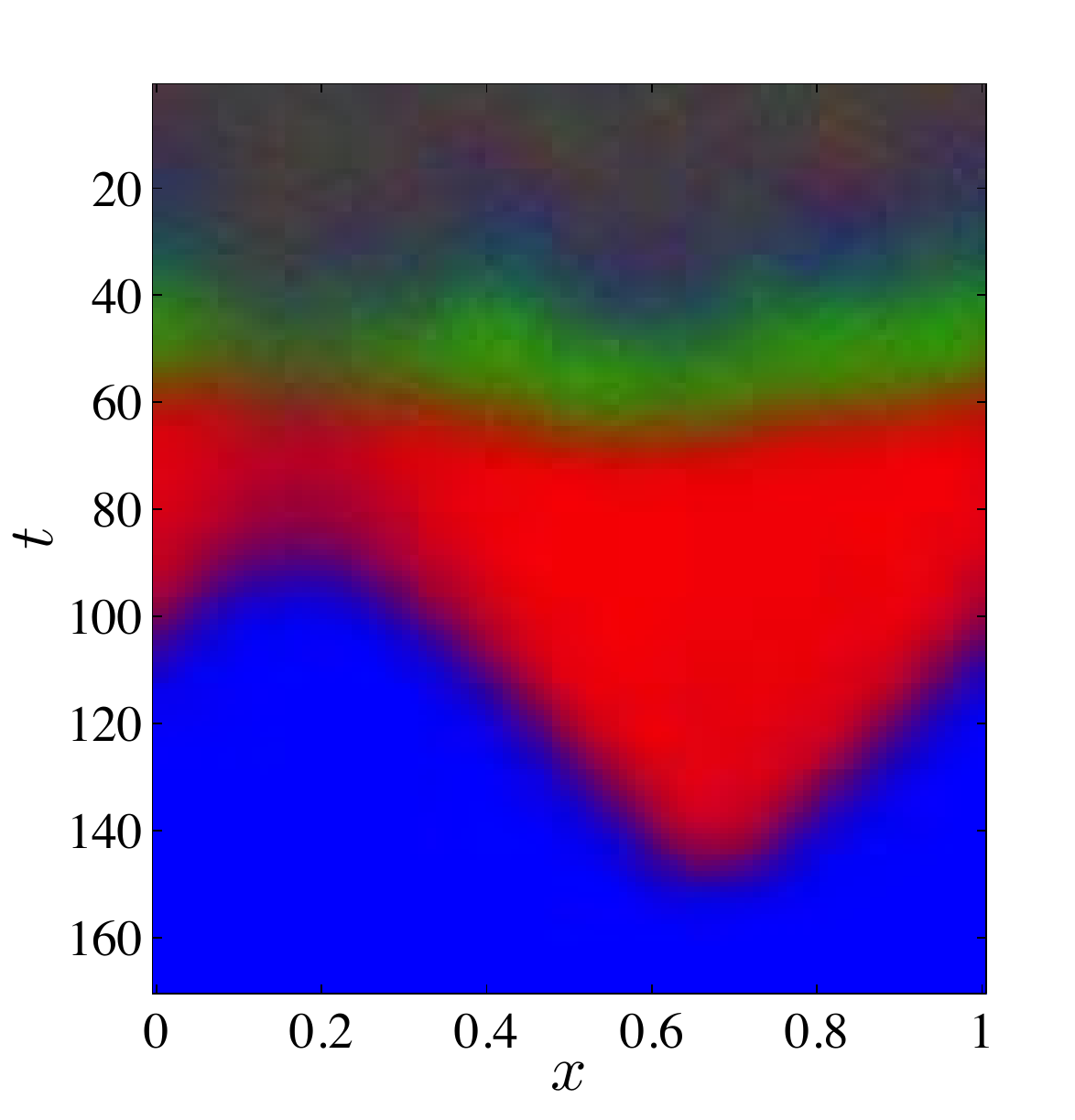}
\includegraphics[width=0.3\textwidth]{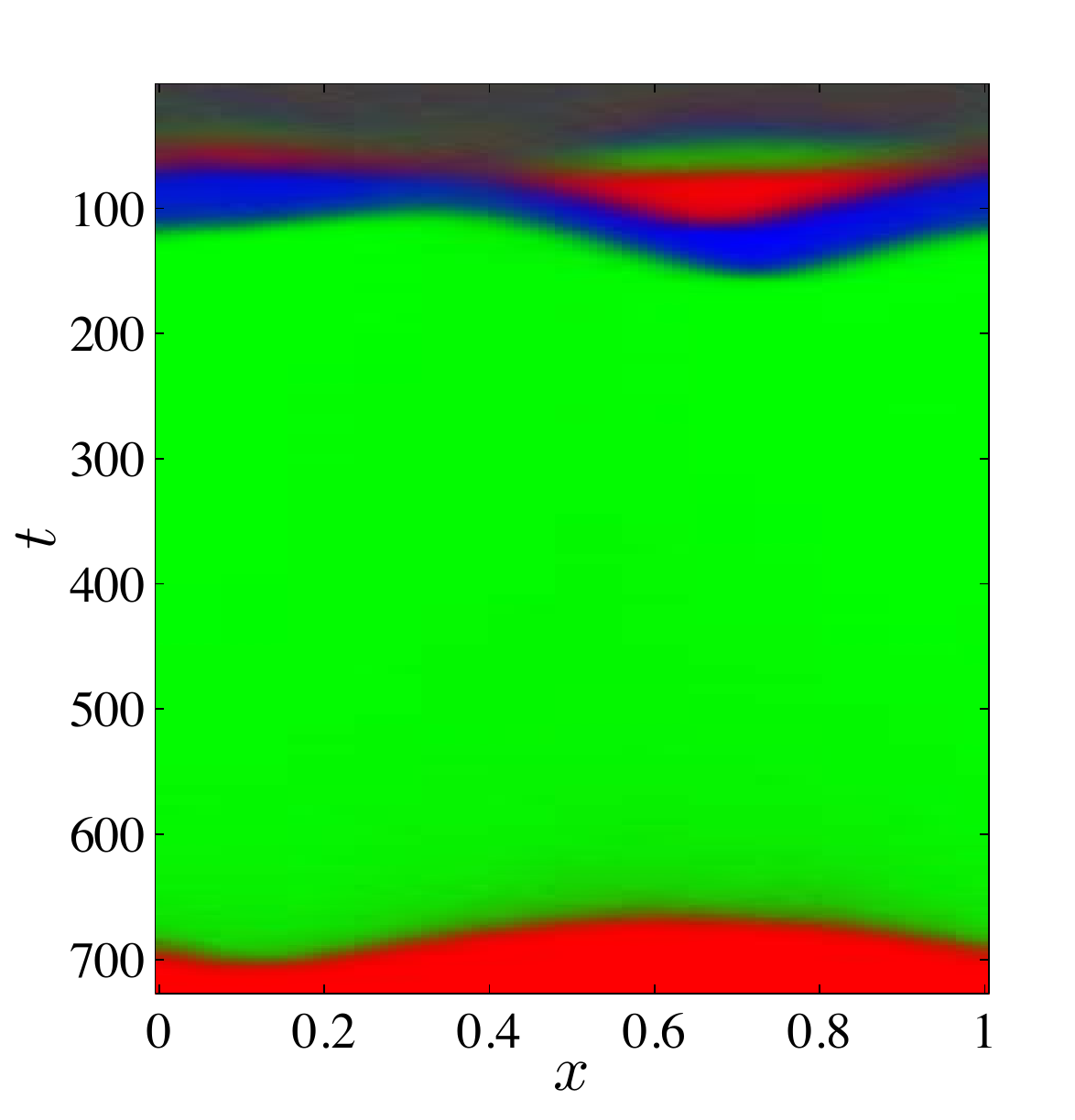}
\includegraphics[width=0.3\textwidth]{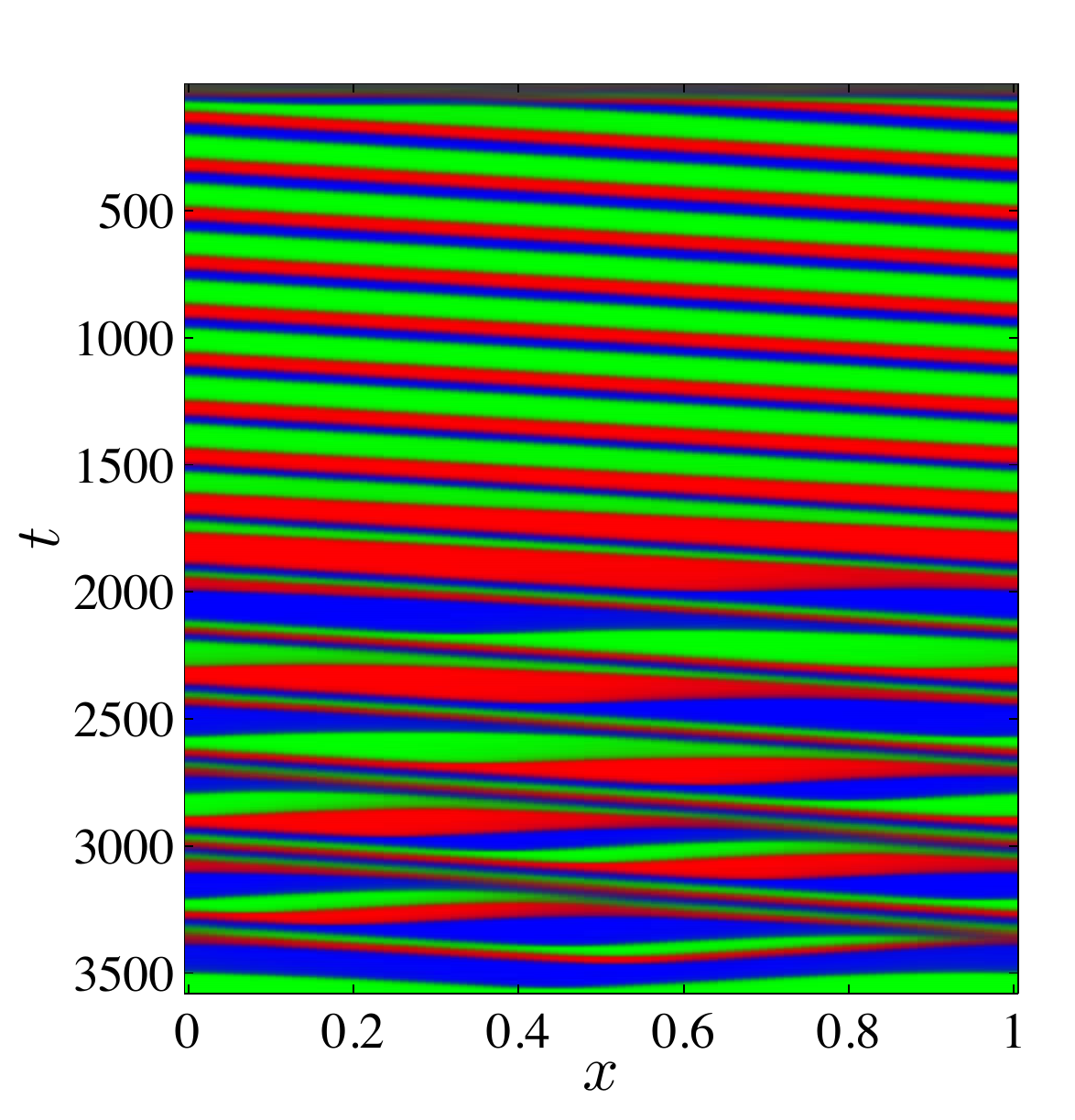}
\caption{(colour online) Spatio-temporal dynamics of three exemplary runs that correspond to the three classes of dynamics ($M=100$, $L=600$): rapid annihilation (left), heteroclinic orbits (center), and propagating waves (right). In the latter case an initially stable wave formation changes periodically at later times, leading to oscillating total densities. Colour encodes the concentrations of the three species (A, red;  B, green; C, blue).\label{fig:2}}
\label{fig:xt}
\end{center}
\end{figure*}

Our simulations reveal three distinct classes of dynamics (Figure~\ref{fig:2}). First, at short time scales stochastic effects lead to the emergence of domains due to coarsening. In this scenario, after a short coarsening process, domains emerge, whose order does not correspond to the rules of cyclic dominance, leading to oppositely moving fronts and hence immediate annihilation. Extinction occurs rapidly in this scenario. The coarsening dynamics to extinction has been extensively studied~\cite{Tainaka-1988,Frachebourg:1996p5895,frachebourg-1996-54}. However, our simulations reveal a much richer dynamics to extinction. Two more processes dominate the dynamics for large times.
Second, we observe situations where the population is almost entirely taken over by a species in cost of a second species, which dies out. A few individuals of the other surviving species are present in the system and, being the dominant one, slowly fixate. This scenario is intimately related  to the heteroclinic orbits of the rate equations~(\ref{eq:re}). The global dynamics moves along the boundary of the invariant manifold of (\ref{eq:re}). Spatial patterns are of minor importance. Third, the system can enter a state of propagating waves of cyclically aligned, uniform domains. These states are only metastable: fluctuating front positions result in domain-annihilation and eventual extinction. For small $D$ this effect was also denoted in \cite{Taeuber2010} and corresponds to the spiral waves found in the two-dimensional model~\cite{Reichenbach2007}. Rare events at the leading edge of the fronts cause the tunneling of domains and oscillating overall species densities.

\begin{figure}[t]
\begin{center}
\subfigure[]{\includegraphics[height=6cm]{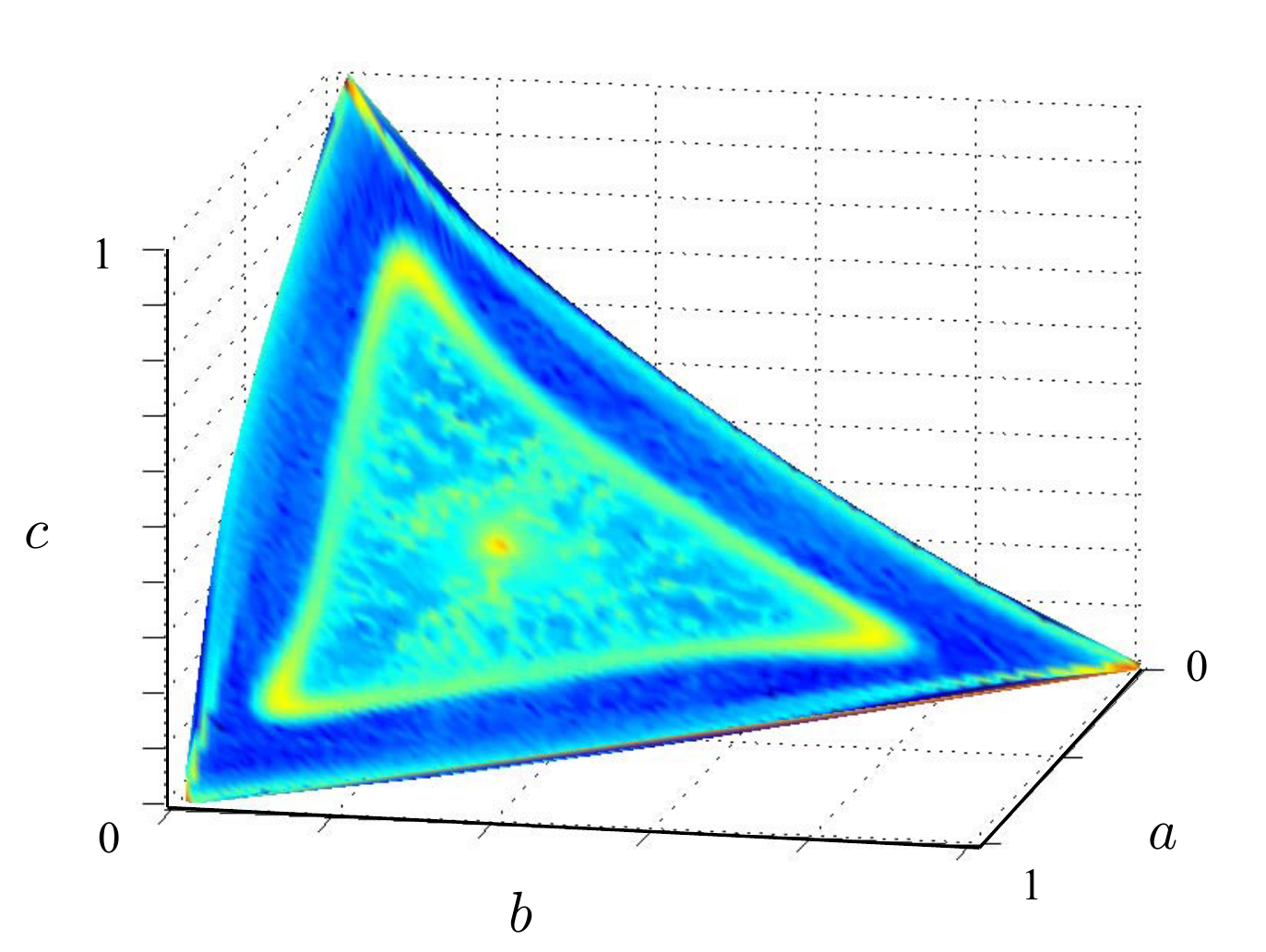}
\label{fig:manifold}}\subfigure[]{
\includegraphics[height=5.4cm]{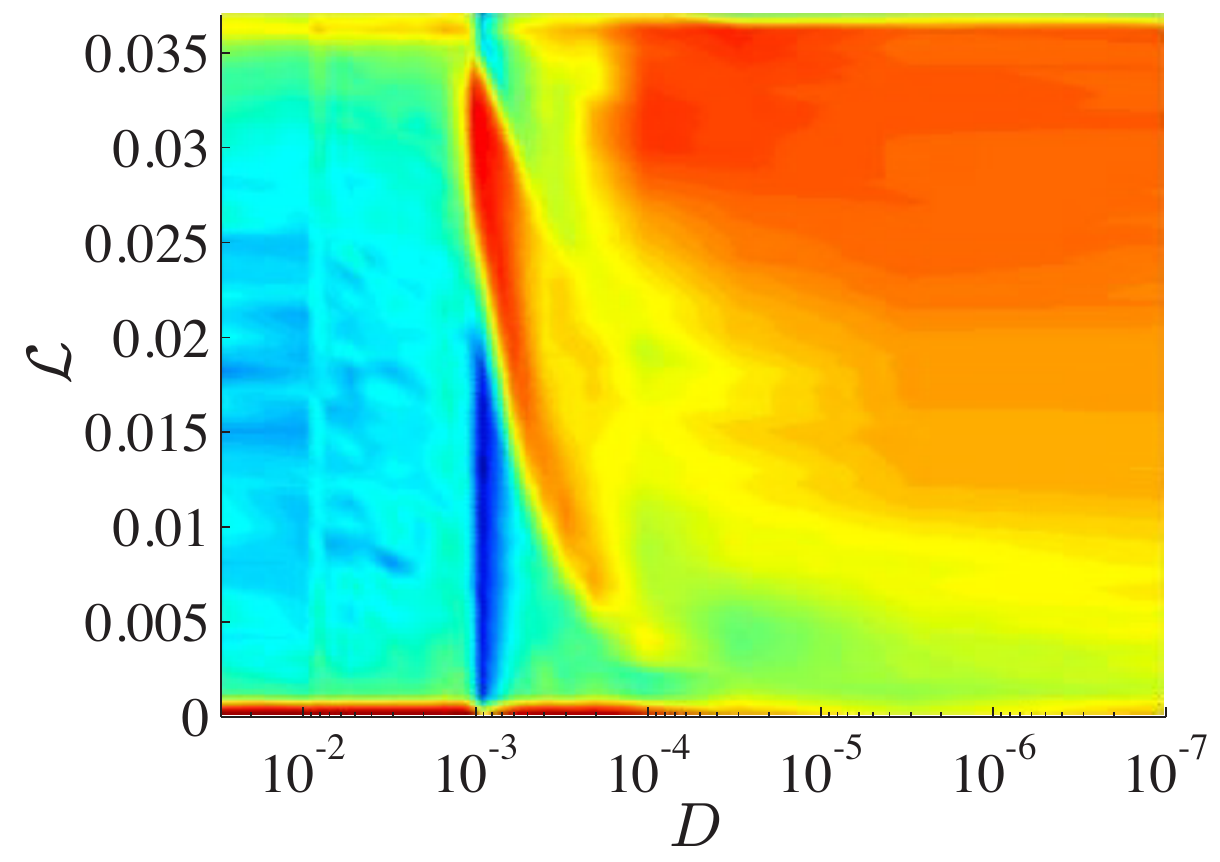}
\label{fig:fed}}
\caption{(colour online) \subref{fig:manifold} Probability of net densities $a,b,c$  for $M=600$, $L=100$, projected onto the invariant manifold of the rate equations~(\ref{eq:re}). Colour encodes  the logarithm of the probability to find the system in the specific state, whereby red denotes the highest, yellow an intermediate, and blue a low probability. Note that the absorbing points themselves have not been included in the statistics. The graph allows to identify the reactive fixed point, an attractor for metastable oscillating waves, and the heteroclinic orbits.
\subref{fig:fed} The Lyapunov function $\mathcal{L}$ provides a measure for the distance of the system's state to the boundary of the simplex. The plot shows the logarithmic probability of net densities for different values of the diffusion constant $D$. Above a critical value of $D$ we find heteroclinic orbits. For very small $D$ we find traveling waves. Below $D_c$ there is a region, where both heteroclinic orbits and traveling waves are present ($M=300$, $L=100$).
}
\end{center}
\end{figure}

Figure~\ref{fig:manifold} provides a concrete picture of the different dynamical processes. The system's state probability is projected onto the invariant manifold of the rate equations~(\ref{eq:re}). The colour signifies the logarithmic probability to find a given net density of species on the manifold before reaching an absorbing fixed point.  We recognize that the system spends considerable time in the vicinity of the boundary, especially near the corners of the simplex, corresponding to the heteroclinic orbits occurring in the second scenario. The \emph{stochastic limit cycle} around the unstable fixed point reflects the oscillating traveling waves from the third scenario (see also, \emph{e.g.},~\cite{PhysRevE.81.066122}).  In the center single trajectories of non-oscillating waves are visible.

The influence of mobility is visualised in Figure \ref{fig:fed}. The Lyapunov function $\mathcal{L}=abc/\rho^3$ characterizes the system's behavior: it is zero at the boundaries and increases monotonically to the unstable fixed point $s^*$. $\mathcal{L}$ therefore provides a measure for the distance of the system's state from the boundaries. The logarithmic probability to find the system at a certain value of $\mathcal{L}$, depending  on the diffusivity $D$, is given in Figure~\ref{fig:fed}. A drastic change in the system's behavior occurs at a critical mobility $D_c\approx 8\cdot 10^{-4}$. Above $D_c$ we observe only heteroclinic orbits, characterized by a high probability to find the system at small values of $\mathcal{L}$. For very small $D$ the system exhibits traveling waves, performing random walks in concentration space. Below $D_c$ both types of dynamics are present. The high probabilities for small values of $\mathcal{L}$ indicate heteroclinic orbits, while the ridge at larger values is caused by oscillating traveling  waves. 

Quantification of the three different dynamical scenarios is feasible through the extinction-time probability, $P(t)$, meaning the probability density that two species go extinct at a certain time $t$. Mathematically it gives the probability distribution function of first-passage times into one of the absorbing fixed points. In our simulations we varied $M$ from 1 to 2800 and set  $D=3\cdot10^{-4}$, \emph{i.e.} in the region, where all types of dynamics arise simultaneously. Figure \ref{fig:logloghist} shows the extinction-time probability distribution for various system sizes. The sharp peak at small times results from the annihilation of oppositely propagating waves as a result of the coarsening process, \emph{i.e.} the first scenario. The functional form of the extinction-time probability distribution for intermediate and large times is determined by the heteroclinic orbits and propagating waves, the second and third types of dynamics. We find an exponentially decaying tail that is,  for large $N$, preceded by a $-3/2$ intermediate  asymptotic power-law interval. The length of this intermediate interval scales linearly with $N$. The plateau or second maximum originates in the short term dynamics of the latter two scenarios.

\section{Semi-phenomenological arguments}

The characteristics of the critical behavior shown in Figure \ref{fig:fed} can be understood through the spatial variant of the rate equations, (\ref{eq:re}). Following Ref.~\cite{reichenbach-2008-254}  the system's dynamics on the invariant manifold can, through a  nonlinear transformation to variables $z_A$ and $z_B$, be recast in terms of the complex Ginzburg-Landau equation
\begin{equation}
\partial_t z = D\nabla^2 +(c_1 -i\omega_0)z - c_2(1+ic_3)|z|^2 z,\label{eq:cgle}
\end{equation}
with $c_1\equiv\frac{\mu\sigma}{2(3\mu +\sigma)}$, $c_2\equiv \frac{\sigma(3\mu +\sigma)(48\mu+11\sigma)}{56\mu(3\mu+2\sigma)}$, and $c_3\equiv \frac{\sqrt{3}(18\mu+5\sigma)}{48\mu+11\sigma}$ \cite{Reichenbach:2007p4097}. The theory of front propagation into unstable states predicts that (\ref{eq:cgle}) always admits traveling waves as stable solutions~\cite{Saarlos2003}. Following a classic treatment of the problem of front-speed selection we obtain their wavelength as $\lambda = -\frac{2\pi c_3\sqrt{D}}{\sqrt{c_1}\left(1-\sqrt{1+c_3^2}\right)}$. At the critical diffusivity $D_c$ the wavelength $\lambda$ exceeds the system size such that the fronts become unstable. From the condition $\lambda=1$ and accounting for the rescaling factor mentioned in \cite{Reichenbach:2007p4097} we obtain $D_c\approx7.6\cdot 10^{-4}$, which is in very good agreement with our numerical results. 

The behaviour of the extinction-time probability distribution can be understood through semi-phenomenological models. In the following we show how such models yield the shape of the extinction-time probability distribution  and its dependence on $N$. In particular, we give an explanation for the scaling behaviour of the power-law interval and the long-time exponential decay. We show that, depending on the system size, either heteroclinic orbits or traveling waves dominate the long-time dynamics. 
\begin{figure}[t]
\centering
\subfigure[]{
\includegraphics[width=0.368\columnwidth]{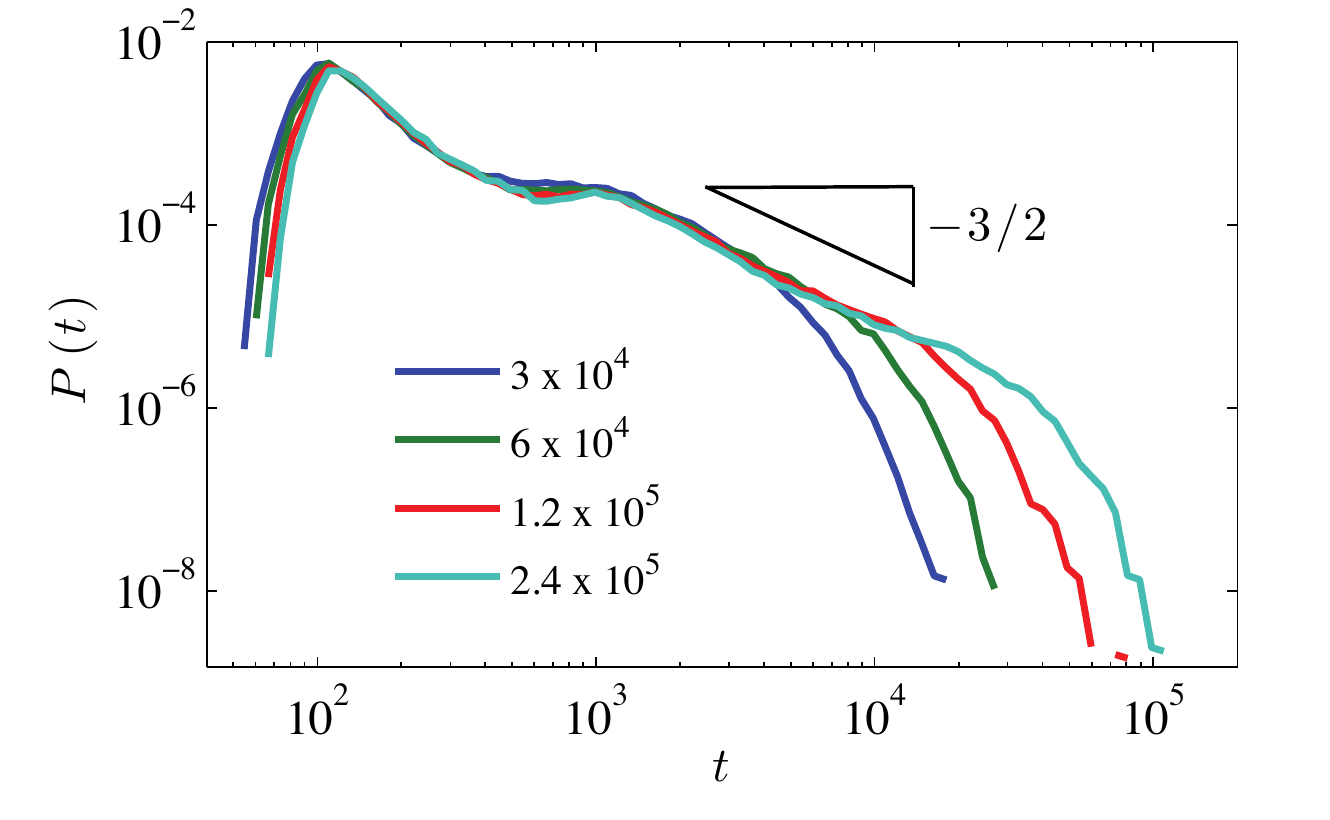}
\label{fig:logloghist}}\subfigure[]{
\includegraphics[width=0.622\columnwidth]{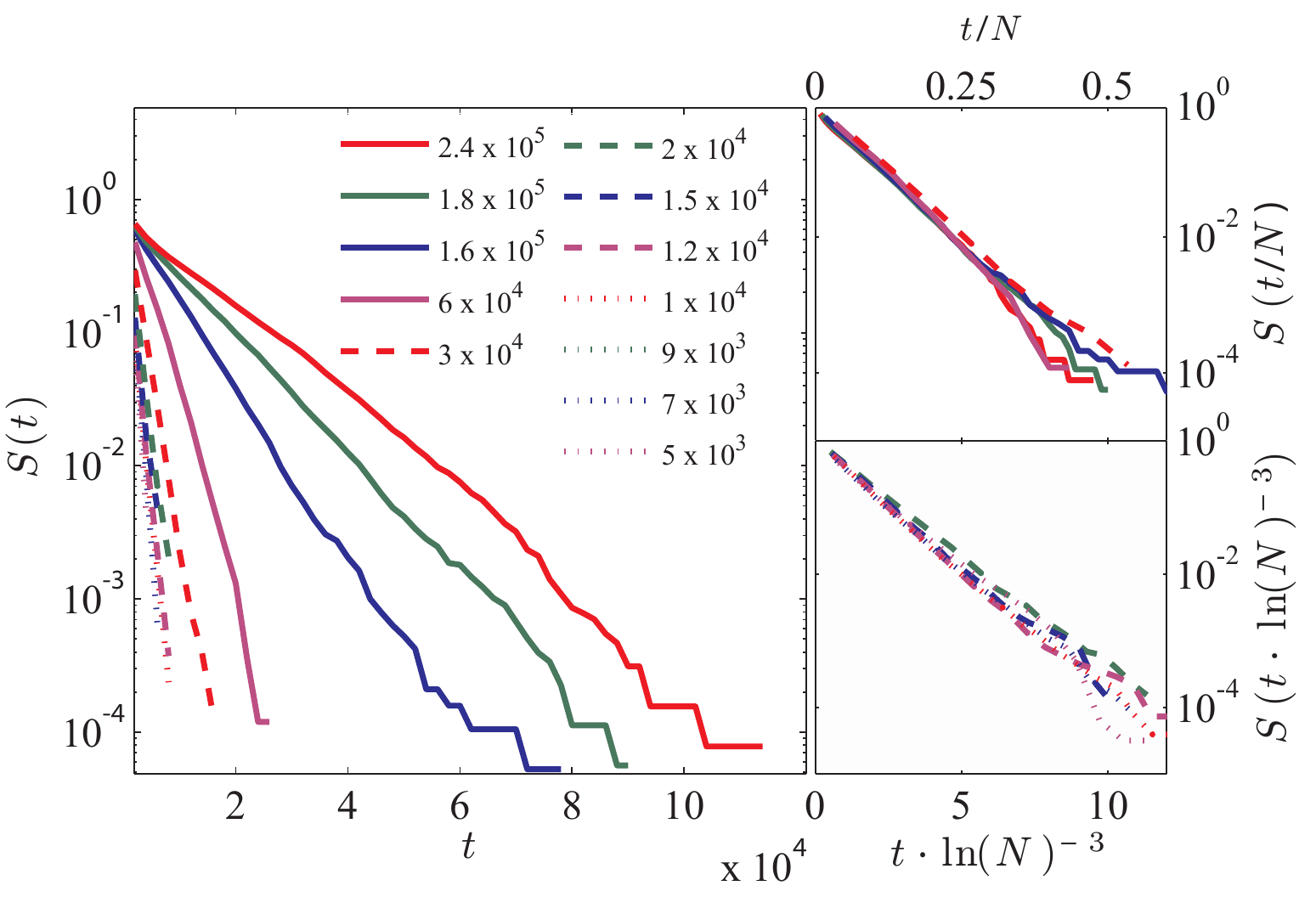}
\label{fig:rescaled}}
\caption{(colour online) \subref{fig:logloghist} Double-logarithmic plot of the extinction-time distribution $P(t)$ for several system sizes. A sharp peak at small times is followed by a second maximum or plateau and by an intermediate $t^{-3/2}$ power law. The length of the power law region scales with $N$. The tail of the distribution decays exponentially. \subref{fig:rescaled} Semi-logarithmic plots of the survival probability $S(t) = 1 -\int_0^t P(t')dt'$ for different $N$. $S(t)$ exhibits the same long-time exponential decay as $P(t)$. With the rescaling $t/N$ for large (top right, $N=30000$ to $N=280000$) and $t\ln(N)^{-3}$ for small systems (bottom right, $N=5000$ to $N=20000$) the exponential tails collapse onto universal curves, in agreement with our analytical predictions.}
\end{figure}

The dynamics of the heteroclinic orbits can be quantified as follows. Consider a small density $a_0\in \mathcal{O}(N^{-1})$ of individuals of species $A$ in a large pool of species $B$ [$b\in \mathcal{O}(1)$]. Due to reproduction of $B$ empty space is as sparse as $A$ individuals are: $1-a-b\in \mathcal{O}(N^{-1})$. Simulations inform us that spatial patterns are not relevant in this scenario. We therefore consider a well-mixed system of size $N$. Three reactions lead to the take-over of the population through the dominating species $A$:
\begin{eqnarray*}
AB\stackrel{\sigma}{\longrightarrow}A\emptyset & \mbox{ at rate } & N\sigma ab\in\mathcal{O}(1),\\
B\emptyset\stackrel{\mu}{\longrightarrow}BB & \mbox{ at rate } & N\mu b(1-a-b)\in\mathcal{O}(1),\\
A\emptyset\stackrel{\mu}{\longrightarrow}AA & \mbox{ at rate } & N\mu a(1-a-b)\in\mathcal{O}(N^{-1}).\end
{eqnarray*}

Here the rates are meant as transitions per unit time. The fast processes are in equilibrium and can be adiabatically eliminated for large $N$, yielding $a=\frac{\mu}{\sigma}(1-a-b)$. Species $A$ occurs at the same density as empty sites times $\mu/\sigma$. A series of reproduction events remains, each with an exponentially distributed waiting time. In the language of stochastic processes this is a pure birth process, studied arising in preferential attachment problems. The extinction-time probability distribution  $P_{h}(t;a_0)$ is thus given by a convolution over exponential functions. By applying the Laplace transform one can show that it can be expressed in closed form as
\begin{equation}
P_{h}(t;a_0)=\sum_{i=0}^{N(1-a_0)}\lambda_{i}e^{-\lambda_{i}t}\prod_{i,j=0,i\neq j}^{N(1-a_0)}\frac{\lambda_
{j}}{\lambda_{j}-\lambda_{i}}\,,
\end{equation}
with rates $\lambda_{i}=N\sigma\left(a_0+i/N\right)^{2}$ \cite{Kannan-1979}. One finally has to marginalize over the probability $p(a_0)$ of starting with an initial density $a_0$ of species $A$ to obtain the extinction-time probability distribution $P_h(t)$ as results from the heteroclinic-orbit dynamics:
\begin{equation}
P_{h}(t)=\sum_{k=1}^{N}P_{h}\left(t;a_0=\frac{k}{N}\right)p\left(a_0=\frac{k}{N}\right).
\label{eq:heteroclinic_orbits}
\end{equation}
For any reasonable $p(a_0)$ the asymptotic behaviour is dominated by the term for the lowest initial concentration $a_0=1/N$ and the lowest reproduction rate $\lambda_{0}$:
$P_h(t)\sim\exp\left(-\sigma t/N\right),\mbox{ for }t\longrightarrow\infty.$
We thus find a $N^{-1}$-dependence of the exponential tail on the system size. For intermediate times we have to take the full convolution and marginalization sums of eq.~(\ref{eq:heteroclinic_orbits}) into account. Numerical evaluation indeed yields a $-3/2$ power-law interval for uniformly distributed $a_0$. The time of crossover between the power-law and the exponential decay scales with $N$.  We therefore find that the second type of dynamics, heteroclinic orbits,  lead to the intermediate power-law regime in the extinction-time probability distribution. The exponential tail of this distribution can either result from heteroclinic orbits or from propagating waves as shown below.

The third type of spatio-temporal dynamics, propagating waves, is metastable. They disappear only through the rare annihilation of neighbouring fronts. By symmetry, the waves move with the same average velocity. For small $D$ the domain interfaces are sharp and therefore interact only on distances that are much smaller than the average domain size. The extinction dynamics in this scenario can therefore be described within an interface picture, where, in a comoving frame, the wave fronts behave as random walkers on a one dimensional lattice with diffusion coefficient $D_f$. For larger $D$ long-range interactions between the interfaces become important, leading to a tunneling of domains. However, within the interface picture this merely corresponds to a relabeling of interfaces and therefore does not influence the extinction dynamics. Numerical simulations validate the assumption of normal diffusion. For  a single series of subsequent $A,B,C$ domains the survival 
probability $S_w(t)$, meaning the probability that all three domains still coexist at time $t$, follows as the survival probability of a single random walker  between absorbing boundaries at distance $l$. The probability distribution $c_w(t,x;x_0,l)$ for the random walker to be at position $x$ and time $t$ when starting at $x_0$ obeys a diffusion equation subject to absorbing boundary conditions. The solution is well known:
\begin{equation}
c_w(t,x;x_0,l)=\sum_{n=1}^{\infty}A_n\sin\left(\frac{n\pi x}{l}\right)e^{-\left(\frac{n\pi}{l}\right)^2 D_{f}t}\,,
\end{equation}
with the coefficients $A_n =\frac{2}{L}\sin\left(\frac{n\pi x_0}{l}\right)$ being determined by the initial condition $c_w(x,t;x_0,l)=\delta(x-x_0)$, see \emph{e.g.}~\cite{Redner2001}. Averaging over space, the initial positions $x_0$ and identically distributed interval lengths $l$ yields the survival probability $S_w(t)$ from the traveling-wave dynamics:
\begin{equation}
S_{w}(t)
=\frac{8}{\pi^{2}}\sum_{m=0}^{\infty}\frac{1}{(2m+1)^{2}}\int_{0}^{1}e^{-(2m+1)^{2}\pi^{2}D_{f}t/l^2}{d} l\,.
\end{equation}
In the asymptotic limit the expression evaluates to
$S_{w}(t)\sim e^{-4D_{f}\pi^{2}t},\mbox{ for }t\longrightarrow\infty$.
The extinction-time probability distribution follows as $P_{w}(t) = -dS_{w}(t)/dt$.
What is the diffusion constant   $D_{f}$ of the domain front? Brunet et al. proposed \cite{Brunet:2006p122} that the diffusion constant for a broad class of stochastic propagating waves depends on $N$ as $D_{f}\sim\ln(N)^{-3}$. Therewith the exponential decay of the extinction-time probability distribution's tail, as resulting from the traveling-wave dynamics, is proportional to $\ln(N)^{-3}$. This result is validated by our numerical simulations, see Figure~\ref{fig:rescaled} bottom right.
%
Heteroclinic orbits and traveling waves both contribute to the asymptotic limit of the net extinction-time probability distribution $P(t)$:
$P(t) \sim P_{h}(t)+P_{w}(t),\mbox{ for }t\longrightarrow\infty$.
Both contributions yield exponential decays at large times, but with different scalings in $N$. For small systems the $\ln(N)^{-3}$ term in $P_w(t)$, resulting from the traveling-wave dynamics, dominates. In contrast,  the $1/N$-decay in $P_h(t)$ resulting from heteroclinic orbits yields the major contribution when $N$ is large. In agreement with these analytical results we indeed find numerically that the exponential decay scales as $\ln(N)^{-3}$ for small $N$ and as $1/N$ for large $N$ (Figure~\ref{fig:rescaled}). Numerically we identified the crossover between both regimes to occur at $N\approx 20,000$. 

\section{Conclusion}
We investigated the spatio-temporal extinction dynamics in a three species stochastic population model with cyclic interactions. While previous work has mainly focused on the coarse graining process that drives the system to extinction we identified two more types of dynamics that are rare but due to their lifetime most important from an evolutionary perspective. The three classes of dynamics, namely rapid annihilation of domains, heteroclinic orbits, and traveling waves are correlated with features of the phase portrait and leave their fingerprints in the extinction-time probability distribution.  The weight of these processes depends on the degree of mixing as well as on the system size. Based on the different dynamical scenarios we provided semi-phenomenological calculations that yield the functional form of this probability distribution and its dependence on the system size. We believe that  our results are of general relevance as we expect a similar phenomenology in other systems described by the complex Ginzburg-Landau equation.


\ack This research was supported by 
the German Excellence Initiative via the program 
`Nanosystems Initiative Munich' and the German Research Foundation via contract FR 850/9-1.
T. R. acknowledges support from the Alexander von Humboldt
Foundation through a fellowship.

\section*{References}
\bibliographystyle{iopart-num}
\bibliography{bib1}

\end{document}